\documentclass[fp,twocolumn]{jpsj3}
\usepackage{txfonts}
\usepackage{color}

\title{Empirical rule of fruit rind fragmentation in muskmelon netting}

\author{Yuri Akiba$^1$, Akari Ishibashi$^2$, Motohiro Sato$^1$, and Hiroyuki Shima$^2$\thanks{hshima@yamanashi.ac.jp}}
\inst{
$^1$Faculty of Engineering, Hokkaido University, Kita 13, Nishi 8, Kita-ku, Sapporo 060-8628, Japan\\
$^2$Department of Environmental Sciences, University of Yamamashi, 4-4-37, Takeda, Kofu, Yamanashi 400-8510, Japan}

\abst{The suberized netting tissue on the surface of muskmelons is a remnant of the rind fragmentation caused by the mechanical imbalance between the expanding flesh meat and the solidified epidermis during growth. The appearance of this netting pattern is an important indicator for assessing the quality of muskmelons; however, very few quantitative studies have been conducted on the geometry of the netting. This study statistically analyzed the areas of fine epidermis fragments surrounded by the net using image analysis and explored a common rule governing the probability distribution of the fragment area. We found that the fragment area distribution follows a universal law by the modified Bessel function, which is consistent with the theoretical argument for the fracture behavior of brittle materials.}

\begin{document}
\maketitle

\section{Introduction}

Melon (Cucumis melo) is a species of cucumber (Cucumis) that has developed into many cultivars \cite{Palenius2008}. Fresh melon meat can be either sweet or bland, with or without a musky aroma; the rind may be netted, ribbed, or smooth. Among the diverse varieties, we will focus on the sweet flesh and musky netted rind type, called muskmelons (or cantaloupe\cite{Trujillo2013}) produced in Japan. Muskmelons produced in Japan are characterized by a uniform and fine reticulate pattern that ornaments their outer surface. We point out here that the reticulate pattern is an interesting topic not only in plant physiology but also in fracture mechanics, as explained below.

\begin{figure}[ttt]
\centering
\includegraphics[width=8.5cm]{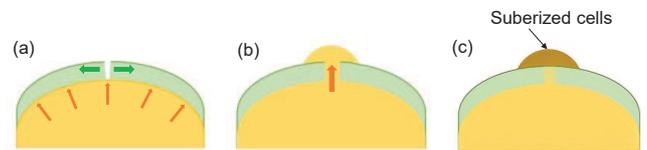}
\caption{
Diagram of the muskmelon netting formation process
(a) Swelling of fruit meat causes a local fracture of the rind.
(b) Flesh cells extrude to the outside of the rind and suberin secreted from the flesh side is deposited on the extruded cell wall. 
(c) Suberin-deposited cell walls thicken and solidify into cork.}
\label{fig01}
\end{figure}

The net of muskmelons is made of solidified secretions that have penetrated the cracks in the rind\cite{Gerchikov2008} (see fig.~\ref{fig01}). In fact, muskmelons do not have a net in the early stage of growth; they grow to the size of chicken eggs about one week after pollination, and at this stage, the outer surfaces are still smooth with no net. Approximately two weeks after pollination, the epidermis becomes hard and enters a hardening period in which the mechanical flexibility of the epidermis is significantly reduced. During this hardening period, the growth rate of the epidermis cannot keep up with the growth rate of the meat because only the inner meat part continues to expand. As a result, many cracks begin to occur hierarchically throughout the epidermis (fig.~\ref{fig01}(a)). Once the surface cracks, a liquid called suberin (a lipophilic polymer) is secreted from the inside, filling the gap in the cracks to prevent moisture loss (fig.~\ref{fig01}(b))\cite{Puthmee2013}. The cork-like solidified secretion in the gap is what we call the net of muskmelons (fig.~\ref{fig01}(c)), which serves as a primary indicator for the non-destructive evaluation of muskmelons.

%--------------
%---------------

It was first pointed out in the early 1950s that cork-like nets that appear on the surface of Cucurbitaceae plants, including muskmelons, originate from cracks on the surface \cite{Meissner1952}. Subsequently, genetic analysis revealed the biochemical process by which suberin secreted into the gap is transformed into cork\cite{KeisermanDiffering2004}. At approximately the same time, it was found that only the netted-rind type of muskmelons contained a large amount of cuticle components in the epidermis, which caused a decrease in the elasticity of the epidermis during the hardening period\cite{KeisermanPeroxidase2004}. More recently, the structure of cell tissue involved in suberization and its biochemical properties have been elucidated by making full use of mass spectrometry and electron microscopy\cite{Cohen2019}. Nevertheless, the scope of these previous studies was limited to either the biochemical attributes or local arrangement of the cell tissue relevant to suberization. Very few examples consider the global geometry of the entire net covering the rind of muskmelons\cite{LLi2012}, despite fundamental interest from the viewpoints of physics and engineering.

From a mechanical perspective, the rind fragmentation of muskmelons can be regarded as hierarchical fracturing of brittle shells under internal pressure-driven expansion. This fracturing process releases deformation energy driven by surface tension, often leaving a polygonal crack network, as observed in organic and inorganic systems, such as powder-water mixtures under desiccation\cite{Nishimoto2007,GoehringSoftMatter2010,Akiba2017,Kitsunezaki2017,Akiba2019,CSTang2021}, volcanic products under cooling\cite{GoehringPNAS2009,Lamur2018,AkibaESS2021,AkibaPRE2021}, animal skin surfaces\cite{ZQin2014}, and tree barks\cite{CSShen2020}, wherein volume contraction or expansion triggers the fragmentation of the material surface. In many cases, whether a particular material breaks or not depends strongly on the measurement conditions and the slight non-uniformity of the mechanical characteristics; therefore, it is not possible to draw a consistent conclusion. In contrast, global fracture patterns that spread throughout a material often follow statistical laws in terms of time or space\cite{CATangPRE2006,ItoPRE2014,ItoJPSJ2014,Halasz2017,Szatmari2021}. Therefore, it can be expected that a hidden statistical law exists in the fracture pattern of muskmelon rind.

In this study, we investigated the geometry of muskmelon netting using image analysis, and scrutinized the presence of statistical laws. We calculated the areas of the rind fragments surrounded by the cork-like substance for multiple specimens and found that the probability distribution of the fragment areas followed a universal curve. Our findings suggest the possibility of automating and objectivizing fruit selection tasks through diagnostic imaging of the rind of muskmelons.

\section{Methods}

%----------------
\subsection{Sample preparation}

We used one of the well-known muskmelon cultivar called ``Raiden'' produced in Hokkaido, northernmost prefecture of Japan, as specimen. Two of the ``Lupia-red" melons and three of the ``Tiara-red" melons, both of which belong to varieties of Raiden melons, were purchased from a supermarket. The five specimens were approximately 14--16 cm in diameter, and they were often slightly vertically long (see Table \ref{table01}). 

\begin{figure}[ttt]
\centering
\includegraphics[width=8.5cm]{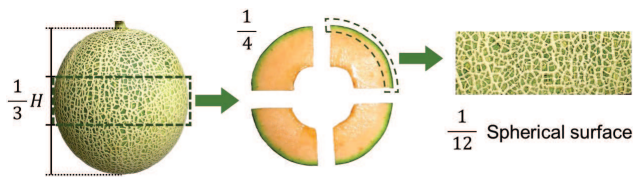}
\caption{Procedure of cutting a muskmelon and drawing the equatorial rind from it.}
\label{fig02}
\end{figure}

\begin{figure}[ttt]
\centering
\includegraphics[width=8.0cm]{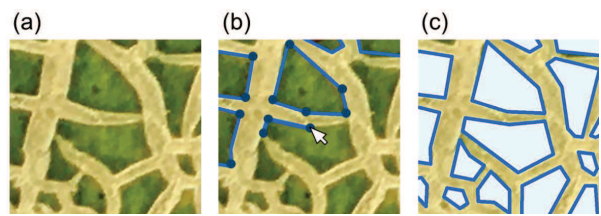}
\caption{Fragment area evaluation by image analysis.
(a) Enlarged photograph of a muskmelon rind containing nets (light green) and fragments (dark green).
(b) Contour line segments (blue lines) and vertices (black dots) were assigned along the boundary of the fragment area.
(c) Polygons (light blue) identified in the fragment area.}
\label{fig03}
\end{figure}

Each specimen was divided vertically into three equal parts (see fig.~\ref{fig02}), and only parts around the equator were extracted. The equatorial part was further divided vertically into four parts, and only the rind was removed. For each of the cut-out rind, dark green fragments surrounded by the light green cork net were collectively photographed from the normal direction of the rind. The obtained photos were used for image analysis; an example of a magnified photograph is shown in fig.~\ref{fig03}(a). We focused only on the surface of the equator, which is far from the fruit stem (top of the specimen) and flower scar (bottom). Thus, it is unnecessary to consider the effect of the presence of fruit stems and flower marks on net shape.

%--------------
%---------------

%--------------
\begin{table}[bbb]
\caption{List of specimen type and dimensions.}
\label{table01}
\begin{center}
\begin{tabular}{cccc}
\hline
\multicolumn{1}{c}{Specimen (cultivar)} 
& \multicolumn{1}{c}{Diameter $D$ (cm)}
& \multicolumn{1}{c}{Height $H$ (cm)}
& \multicolumn{1}{c}{$D/H$} \\
\hline
a (Tiara-red) & 14.3 & 15.7 & 0.91 \\
b (Tiara-red) & 15.0 & 15.0 & 1.00 \\
c (Tiara-red) & 14.6 & 15.5 & 0.94 \\
d (Lupia-red) & 16.0 & 17.0 & 0.94 \\
e (Lupia-red) & 15.0 & 18.0 & 0.83 \\
\hline
\end{tabular}
\end{center}
\end{table}
%--------------

It is noteworthy that we dealt only with commercially available muskmelons; this implies that  only individuals exhibiting such a uniform and fine mesh pattern that can pass the selection criteria in fruit sorting have been the subject of research. As well, it is worth mentioning that the fragmented rind we examined was stationary in the sense that the crack network did not develop further over time. In fact, after the fruit of the melon is detached from the branch, the water supply to the fruit ceases; as a result, the flesh will stop swelling and no new cracks will develop in the rind.

%--------------
%--------------

%----------------
\subsection{Image analysis}

To examine the geometric area of the rind fragments and their probability distribution, the photos obtained were analyzed using an image processing software (ArcGIS, Esri).

First, the Cartesian coordinates of the actual length scale were assigned to the image data. Next, to extract the fragment areas enclosed by the net tissue from the image, the outline of the fragments was traced, and line data were created (see fig.~\ref{fig03}(b)). The contour was specified manually because the color difference between the net (light green) and fragments (dark green) appears rather ambiguous, making it technically difficult for ArcGIS to distinguish them. Finally, the line data were converted into polygon data (fig.~\ref{fig03}(c)) and evaluated the fragment areas. This procedure was performed on all photographs obtained from the five specimens to calculate the frequency distribution of the fragment areas.

%--------------
\begin{table}[bbb]
\caption{
List of the normalization constants $s_0$.
The five values on the left (labeled by a-e) are the values of $s_0$ calculated by dealing with the data of $s$ for each specimen separately.
The rightmost value (labeled by All) is that obtained by dealing with the data for all the five specimens collectively.
}
\label{table02}
\begin{center}
\begin{tabular}{ccccccc}
\hline
  \multicolumn{1}{c}{Specimen}
& \multicolumn{1}{c}{a}
& \multicolumn{1}{c}{b}
& \multicolumn{1}{c}{c}
& \multicolumn{1}{c}{d}
& \multicolumn{1}{c}{e}
& \multicolumn{1}{c}{All} \\
\hline
$s_0$ [mm$^2$] & 
4.9 & 2.6 & 4.6 & 6.0 & 4.3 & 3.8 \\
\hline
\end{tabular}
\end{center}
\end{table}
%--------------

%---------------
\section{Results}

%----------------
\subsection{Occurrence frequency of fragment area}

\begin{figure}[ttt]
\centering
\includegraphics[width=6.5cm]{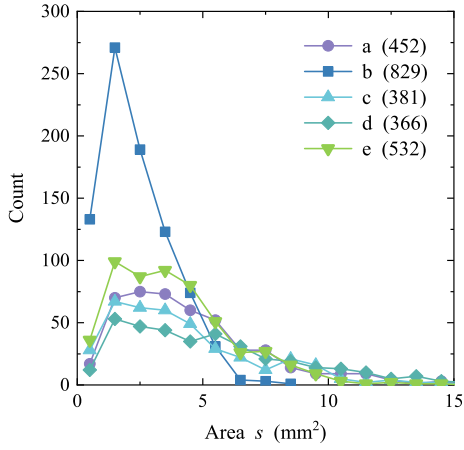}
\caption{Occurrence frequency of the fragment area $s$ for five different muskmelon specimens.
The legend indicates the number of fragment samples for each muskmelon specimen.}
\label{fig04}
\end{figure}

Figure \ref{fig04} shows the frequency of occurrence of fragment area $s$ over the five different specimens listed in Table \ref{table01}. The five measurement datasets commonly show one wide peak slanted to the left with a fat tail at large $s$, while the peak position and peak height vary greatly between specimens. The total frequency of occurrence also varied from specimen to specimen, as presented in the legend. These observations indicate that the net shape of muskmelons is highly dependent on the growth conditions and exhibits large fluctuations. That is, there is no common rule for the frequency of the fragment area $s$ itself.

In contrast to the above results, if we replace the horizontal axis from $s$ with a normalized one, a hidden law becomes apparent. Figure \ref{fig05} shows the probability distribution of the measurement data $s/s_0$; here, the value of $s_0$ was selected such that the expected value of $s/s_0$ for each specimen was equal to unity. The calculated values of $s_0$ are listed in Table \ref{table02}. It is clearly seen from fig.~\ref{fig05} that the probability distributions of $s/s_0$ derived from five different specimens show very similar curves with slight variation. This was the first main result of our study.

By definition, $s_0$ is the expected value of the fragmentation area for each specimen; the finer (or coarser) the netting pattern in the specimen's rind, the smaller (larger) the value of $s_0$ in principle. Still it is nontrivial to formulate the value of $s_0$ as a function of other physical factors, as it is greatly affected by individual growth processes and stochastic environmental changes. At least from the comparison of Tables \ref{table01} and \ref{table02}, it can be inferred that $s_0$ does not correlate with the size or aspect ratio of melon individuals. Despite such the strong specimen dependence of $s_0$, the probability distribution of $s/s_0$ turned out to obey a specimen-independent curve as demonstrated in fig.~\ref{fig05}.

%----------------
\subsection{Two choices for fitting curve}

To account for the common behavior of the probability distribution of $s/s_0$, we chose two specific classes of fitting curves. The first is represented by a function of
\begin{equation}
P_{\rm w}(s) =
\frac{2k}{s_0^{(\rm w)}} \left( \frac{s}{s_0^{(\rm w)}} \right)^{k-1}
K_0 \left[ 2 \left( \frac{s}{s_0^{(\rm w)}} \right)^{k/2} \right],
\label{eq_111}
\end{equation}
where $k$ is the fitting parameter and $K_0(u)$ is called the zero-th order modified Bessel function of the second kind, which monotonically decreases with $u$. In eq.~(\ref{eq_111}), the value of $s_0^{(\rm w)}$ is set to 
\begin{equation}
s_0^{(\rm w)} = \left[ \Gamma\left( 1 +\frac{1}{k} \right) \right]^{-2},
\label{eq_004}
\end{equation}
where $\Gamma(z)=\int_0^\infty u^{z-1} e^{-u} du$ denotes the gamma function.
The setting of eq.~(\ref{eq_004}) ensures that the normalization condition
\begin{equation}
\int_0^{\infty} s P_{\rm w}(s) ds = 1,
\label{eq_004x}
\end{equation}
is satisfied for any value of $k$; see Discussion in Sec.~\ref{subsec32}. 
The other fitting curve is represented by the following:
\begin{equation}
P_{\rm g}(s) =
\frac{2}{s_0^{(\rm g)}} \left( \frac{k}{\Gamma(k)} \right)^2 \left( \frac{k^2 s}{s_0^{(\rm g)}} \right)^{k-1}
K_0 \left( 2k \sqrt{\frac{s}{s_0^{(\rm g)}}} \right),
\label{eq_222}
\end{equation}
where $s_0^{(\rm g)}=1$ is fixed, such that the normalization condition similar to eq.~(\ref{eq_004x}) is satisfied. The optimized values of the fitting parameter $k$ are generally different for $P_{\rm w}(s)$ and $P_{\rm g}(s)$.

%--------------
%--------------

\begin{figure}[ttt]
\centering
\includegraphics[width=6.4cm]{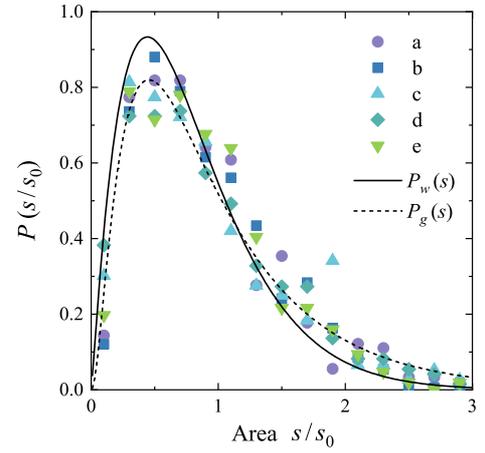}
\caption{Normalized probability distribution of the normalized fragment area $s/s_0$.
The solid and dashed curves defined by eqs.~(\ref{eq_111}) and (\ref{eq_222}) are the optimized fitting curves for all data points, with settings of $k=2.4$ and $k=3.8$.}
\label{fig05}
\end{figure}

Figure \ref{fig05} shows the optimal fitting curves described by eqs.~(\ref{eq_111}) and (\ref{eq_222}). The optimized values of $k$ for all data points were $k=2.4$ and $k=3.8$ for the curves of $P_{\rm w}(s)$ and $P_{\rm g}(s)$, respectively. Both curves are in good agreement with the measured data. We confirmed that these fittings work well even if the data points are separated by specimens, and Table \ref{table03} presents the list of optimized $k$ values when the data obtained from five individual specimens are separately fitted by eqs.~(\ref{eq_111}) and (\ref{eq_222}). The fluctuations in the $k$ value for the five specimens were within 13 \% and 26 \% for the $P_w$- and $P_g$-based fits, respectively.
Minor fluctuations in $k$ support the validity of our choice of fitting distribution functions for muskmelons of various sizes. The discovery of two appropriate fitting distribution functions for $s/s_0$ of muskmelon nets is the second main finding of this study.

%--------------
%--------------

%--------------
\begin{table}[bbb]
\caption{List of the optimized values $k$ as fitting parameters.}
\label{table03}
\begin{center}
\begin{tabular}{ccccccc}
\hline
  \multicolumn{1}{c}{Specimen}
& \multicolumn{1}{c}{a}
& \multicolumn{1}{c}{b}
& \multicolumn{1}{c}{c}
& \multicolumn{1}{c}{d}
& \multicolumn{1}{c}{e}
& \multicolumn{1}{c}{All} \\
\hline
$P_{\rm w}(s)$& 
2.6 & 2.6 & 2.3 & 2.3 & 2.6 & 2.4 \\
$P_{\rm g}(s)$ &
4.4 & 4.5 & 3.6 & 3.5 & 4.3 & 3.8 \\
\hline
\end{tabular}
\end{center}
\end{table}
%--------------

%--------------
\begin{table}[bbb]
\caption{List of the minimum SSE obtained when fitting eqs.~(\ref{eq_111}) and (\ref{eq_222}) to the measurement data of $P(s)$
by the optimized fitting parameter $k$.}
\label{table04}
\begin{center}
\begin{tabular}{ccccccc}
\hline
  \multicolumn{1}{c}{Specimen}
& \multicolumn{1}{c}{a}
& \multicolumn{1}{c}{b}
& \multicolumn{1}{c}{c}
& \multicolumn{1}{c}{d}
& \multicolumn{1}{c}{e}
& \multicolumn{1}{c}{All} \\
\hline
$P_{\rm w}(s)$& 
0.0598 & 0.0418 & 0.0714 & 0.0217 & 0.0478 & 0.0550 \\
$P_{\rm g}(s)$ &
0.0543 & 0.0304 & 0.0583 & 0.0259 & 0.0592 & 0.0220 \\
\hline
\end{tabular}
\end{center}
\end{table}
%--------------

%----------------
\subsection{Comparison of the goodness of fittings}

In actual calculations, the goodness of fitting was examined using the value of the sum of squares of error (SSE). 
The optimal value of $k$ was determined such that it minimizes the SSE between the measurement data of $P(s)$
and the values either of $P_w(s)$ or $P_g(s)$. The numeric results of SSE are listed in Table \ref{table04}.
It follows from Table \ref{table04} that the curve of $P_g(s)$ seems to be better for the fitting 
if all the data obtained from the five specimens are collectively examined.
However, if we examine each sample individually, we cannot determine which curve is better for fitting.
This issue will be revisited in Sec.~\ref{sec_open}.

%----------------
\section{Discussion}

%----------------
\subsection{Derivation of $K_0$-based fitting functions}\label{subsec32}

The reason why our experimental data of the fragment area $s$ obey the distribution functions $P_{\rm w}(s)$ and $P_{\rm g}(s)$ defined by eqs.~(\ref{eq_111}) and (\ref{eq_222}), respectively, can be explained as follows.

In the field of statistical physics, it is known that\cite{CLuPRE2002,CarmonaPRE2008,Paluszny2016} the size distribution of fragments produced by the fracture of brittle materials often follows a unique probability distribution function, which may belong to the Weibull or gamma distribution. A typical example is the fracture spacing in one-dimensional ring-shaped brittle materials under radial expansion\cite{Mott1947}. It was proved that the probability density function of the fracture spacing in the circumferential $(x)$ direction is approximately represented by either of the following two functions\cite{Grady1981,Grady2008}; the one is Weibull distribution written by
\begin{equation}
w(x) = \frac{k}{x_0} \left( \frac{x}{x_0} \right)^{k-1} e^{-(x/x_0)^k},
\label{eq_555w}
\end{equation}
satisfying $\int_0^{\infty} w(x)dx=1$, and the gamma distribution is defined as
\begin{equation}
g(x) = \frac{1}{x_0} \frac{k}{\Gamma(k)} \left( \frac{kx}{x_0} \right)^{k-1} e^{- kx/x_0},
\label{eq_555g}
\end{equation}
that satisfies $\int_0^{\infty} g(x)dx=1$. Both distributions contain two parameters, $x_0$ and $k$, and in actual fracture phenomena, the values of the two are dependent on the material properties and experimental conditions. For each distribution, the expected value of $x$ is expressed as
\begin{equation}
\langle x\rangle_{\rm w} = \int_0^\infty x w(x) dx = x_0 \Gamma\left( 1+\frac{1}{k}\right),
\label{eq_037}
\end{equation}
and 
\begin{equation}
\langle x\rangle_{\rm g} = \int_0^\infty x g(x) dx = x_0.
\label{eq_038}
\end{equation}

We now hypothesize that the analysis of the fracture distribution of one-dimensional rings can be extended to explain the collapse of the expanding spherical shell segments under stretching in both the latitude $(x)$ and longitude $(y)$ directions. Suppose the rind of a muskmelon during growth can be regarded as an expanding spherical shell subject to internal pressure. We constantly observed a small portion of the shell until the fracture. We also assumed that the fracture spacing within the portion follows the distribution in either of eq.(\ref{eq_555w}) or (\ref{eq_555g}) in the $x$ and $y$ directions. Once one of the two is chosen, we can theoretically estimate the fragment area distribution within a portion of the shell in the following manner.

If we choose the former type of distribution $w$, then the two-dimensional statistical partitioning $f(x,y)$ is given by the product of the two distributions, $f(x,y) = w(x)w(y)$, which reads
\begin{equation}
f(x,y) = \frac{k^2}{x_0 y_0} \left( \frac{xy}{x_0 y_0} \right)^{k-1} 
\exp \left[ -\left( \frac{x}{x_0}\right)^k -\left( \frac{y}{y_0}\right)^k \right].
\label{eq_040}
\end{equation}
We then transform the variables from $x,y$ to $s,q$, defined as
\begin{equation}
s = xy, \quad q=\frac{x}{y},
\label{eq_045}
\end{equation}
to obtain the probability density function $h(s,q)$ in terms of $s,q$ as follows:
\begin{eqnarray}
h(s,q) 
&=&\!
f(x(s,q),y(s,q)) \left| \frac{\partial (x,y)}{\partial (s,q)} \right| \nonumber \\
&=&\!
\frac{k^2}{2 (x_0 y_0)^k} \frac{s^{k-1}}{q} 
\exp \left[ 
-\left(\frac{\sqrt{sq}}{x_0} \right)^k 
-\left(\frac{1}{y_0} \sqrt{\frac{s}{q}} \right)^k
\right].\qquad
\label{eq_045}
\end{eqnarray}
Consequently, the fragment area distribution $P_{\rm w}(s)$, based on the fracture spacing $w(x)$, is given by
\begin{equation}
P_{\rm w}(s) = \int_0^\infty h(s,q) dq.
\label{eq_048}
\end{equation}
To proceed with the calculation, we further transform the variable from $q$ to $\eta$ using the following relation:
\begin{equation}
q = \frac{x_0}{y_0} \exp\left( \frac{2\eta}{k}\right).
\end{equation}
Substituting it to eq.~(\ref{eq_048}), we obtain
\begin{eqnarray}
P_{\rm w}(s)
&=& 
\int_{-\infty}^{\infty} g(s,q(\eta)) \frac{dq}{d\eta} d\eta \nonumber \\
&=&
\frac{2k}{s_0} \left( \frac{s}{s_0} \right)^{k-1}
\int_0^\infty \exp\left\{ -2 \left( \frac{s}{s_0} \right)^{k/2} \cosh \eta \right\}d\eta,\qquad
\label{eq_049}
\end{eqnarray}
with $s_0 = x_0 y_0$. The integration in eq.~(\ref{eq_049}) is equivalent to the integral representation of the modified Bessel function of the second kind, $K_0$. Eventually, we arrive at the desired distribution function of eq.~(\ref{eq_111}).

In the case of choosing another distribution, $g$, for the fracture spacing, parallel discussion similar to above applies; we consider the product $f(x,y) = g(x)g(y)$, which leads to the associated $h(s,q)$ and $P_{\rm g}(s) = \int_0^\infty h(s,q) dq$. Through variable transformation via $q = (x_0/y_0) e^{2\eta}$, we obtain the fragment area distribution $P_{\rm g}(s)$ based on the fracture spacing $g(x)$, as eq.~(\ref{eq_222}).

%----------------
\subsection{Probability distribution $p(\ell)$ of crack spacings $\ell$}

Our discussion in the previous subsection was based on the following two primary assumptions:
i) the one-dimensional fracture spacings obey either of the Weibull distributions $w(x)$ or gamma distributions $g(x)$, and ii) the two-dimensional statistical partitioning $f(x,y)$ is equal to the product of the two one-dimensional fracture-spacing distributions (e.g., $w(x)$ and $w(y)$). Considering this discussion, it would be interesting to examine the validity of the assumption i) on the muskmelon rind fragmentation. To this aim, we evaluated the probability distribution of the fracture spacings in circumferential direction using the same photos explained in Section 2.2.

\begin{figure}[ttt]
\centering
\includegraphics[width=8.4cm]{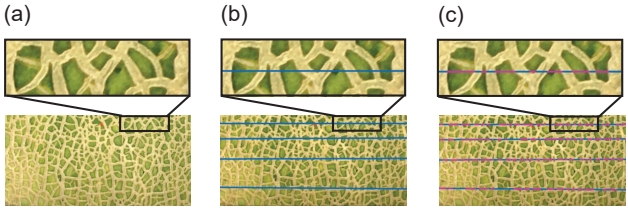}
\caption{Schematics of the fracture spacing analysis. (a) Original photo. (b) Parallel horizontal lines (colored in blue) assigned to the image. (c) A set of line segments (highlighted in magenta) that overlaps with the fragment area of the muskmelon rind (colored in dark green)..}
\label{fig06add}
\end{figure}

Figure \ref{fig06add} illustrates the procedure of analyzing the fracture spacings. Given an original image data (fig.~\ref{fig06add}(a)), we assigned to the image several horizontal lines colored in blue in fig.~\ref{fig06add}(b). The direction of those lines is set to be parallel to the equator of the muskmelon, while the spacing between these lines was set randomly. From the continuous point sequence that composes these lines, we extracted a set of point sequences that overlaps with the fragment area of the muskmelon rind; the set of point sequences is highlighted by short line segments colored in magenta in fig.~\ref{fig06add}(c). We then evaluate the length of every line segment, designated by $\ell$, and computed the probability distribution of $\ell$ for each of the five specimens. When plotting the results, the variable $\ell$ is divided by the expected value $\ell_0$ for each specimen for normalization.

\begin{figure}[ttt]
\centering
\includegraphics[width=6.2cm]{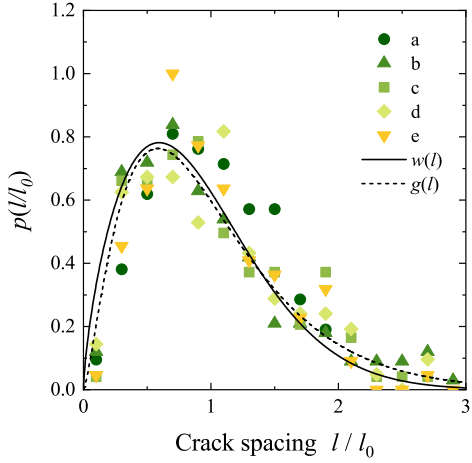}
\caption{Normalized probability distribution of the crack spacings $\ell/\ell_0$.
The solid and dashed curves defined by eqs.~(\ref{eq_555w}) and (\ref{eq_555g}) are 
the optimized fitting curves for all data points, with settings of $k=1.7$ and $k=2.4$.}
\label{fig07}
\end{figure}

Figure \ref{fig07} shows the normalized probability distribution of $\ell$ that was experimentally obtained. The fitting curves to all the data points, based on the Weibull distribution (see eq.~(\ref{eq_555w})) and the gamma distribution (eq.~(\ref{eq_555g})) are also plotted by the solid and dashed curves, respectively. The optimal values of the fitting parameter $k$ in the two distribution functions are summarized in Table \ref{table05}. It is clearly observed from fig.~\ref{fig07} that both the two curves fit well to the measured data, implying that the fracture spacing along the horizontal perimeter of the specimen obeys statistically either of the two probability distribution. This result supports the validity of our discussion in the previous subsection. 

%--------------
%--------------

%--------------
%--------------

%--------------
\begin{table}[bbb]
\caption{ List of the optimized values $k$ for the fitting of crack-spacing probability distributions shown in fig.~\ref{fig07}.}
\label{table05}
\begin{center}
\begin{tabular}{ccccccc}
\hline
  \multicolumn{1}{c}{Specimen}
& \multicolumn{1}{c}{a}
& \multicolumn{1}{c}{b}
& \multicolumn{1}{c}{c}
& \multicolumn{1}{c}{d}
& \multicolumn{1}{c}{e}
& \multicolumn{1}{c}{All} \\
\hline
$w(\ell)$& 
2.1 & 1.8 & 1.8 & 1.8 & 2.1 & 1.7 \\
$g(\ell)$ &
3.7 & 2.4 & 2.6 & 2.5 & 3.5 & 2.4 \\
\hline
\end{tabular}
\end{center}
\end{table}
%--------------

%----------------
\subsection{Physical meaning of the optimal fitting parameter $k$}

%--------------
\begin{figure}[ttt]
\centering
\includegraphics[width=4.0cm]{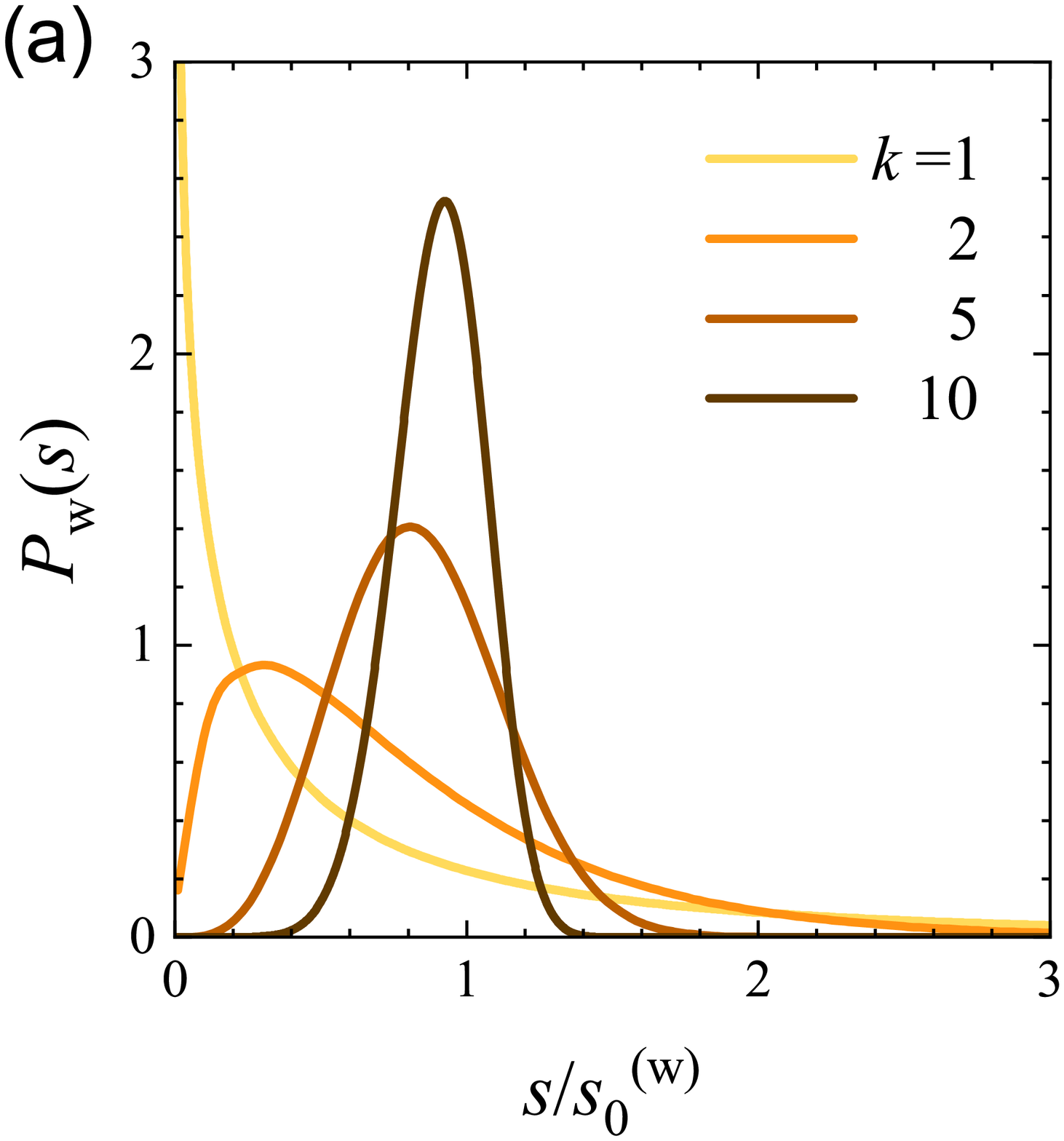}
\includegraphics[width=4.0cm]{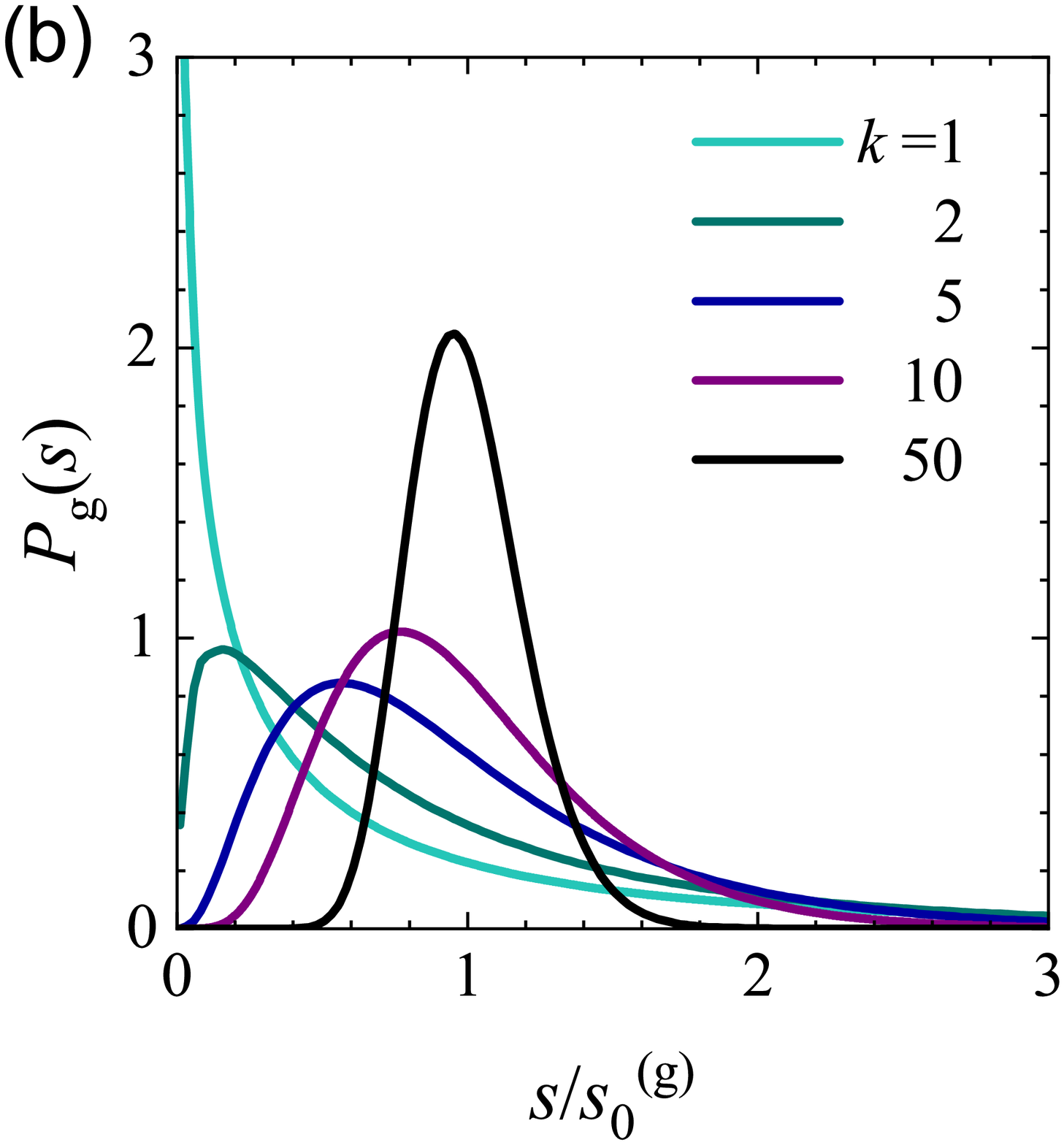}
\caption{Variations in the two distribution curves with increasing $k$: (a) $P_w(s)$ and (b) $P_g(s)$.}
\label{fig06}
\end{figure}
%---------------

In Section 3, we have used as the fitting parameter the variable $k$ contained in the functions of $P_w(s)$ and $P_g(s)$, defined by eqs.~(\ref{eq_111}) and (\ref{eq_222}), respectively. From physical (or geometric) viewpoints, the magnitude of the parameter $k$ represents the uniformity of the fragment area over the specimen's rind. 

The above physical interpretation of $k$ can be explained by fig.~\ref{fig06}, which shows the gradual change of the curves $P_w(s)$ and $P_g(s)$ with increasing $k$. The two curves show an upward peak and a heavy tail for $k= 2$ (or more), whereas they are monotonically downward-sloping for $k=1$ (or less). An important observation is that with increasing $k$, the peak height increases and the peak position approaches $s/s_0^{(\alpha)}=1$  ($\alpha={\rm w, g}$). This result means that the larger the value of $k$, the more uniform the fragment area and the closer to the average value (i.e., $s_0^{(\alpha)}$).

Considering the growth process of the muskmelon, the uniformity of the fragment area is partly determined by the uniformity of the internal pressure that exerts from the meat to the rind. If the growth rate of the meat is spatially uniform and if the melon shape is ideally spherical, then the meat expansion results in uniform inner pressure exerted on the rind and then uniform fragmentation of the rind characterized by the limit of $k\to \infty$. In reality, however, the growth rate of the meat should fluctuate spatially and the shape of the melon becomes slightly elongated ellipsoid. Our measured values of $k$ are thought to indicate the degree of deviation from this ideal uniform state.

%----------------
\subsection{Open question: Which is better: $P_w(s)$ or $P_g(s)$?}\label{sec_open}

We have demonstrated that the two specific functions, $P_w(s)$ and $P_g(s)$, describe well the measured data of the two-dimensional fragment area distribution in muskmelons. This finding may pose a question under what physical conditions is $P_w(s)$ or $P_g(s)$ more appropriate to describe the rind fragmentation. From a general point of view (not limited to muskmelon rind), the Weibull distribution is known to work well when a material is brittle and homogeneous\cite{CLu2002,Danzer2007}; in addition, the independence between different cracks and the weakest-link hypothesis (i.e., the weakest defect causes failure of the whole structure) should be satisfied\cite{Peterlik2001}. Hence, $P_w(s)$ will be more appropriate as a fitting curve in the present work, if every muskmelon rind satisfies the conditions mentioned above. However, it must be rare for actual materials to meet all the conditions, both in specific (i.e., muskmelon rind) and general cases. In order to deal with real experimental data, therefore, various statistical models different from the Weibull type have been thus far proposed in the community of fracture mechanics\cite{Gorjan2012,WSLei2020}; the gamma distribution can be categorized as one of them\cite{Wakabayashi2009}. Still yet, the question as to which probability distribution is the most optimal under what physical conditions has been a long-standing problem, even from a general perspective, and has remained to be clarified at present. 

Returning to the issue specific to muskmelon rind, the authors do not enter into the question of which curve, $P_w(s)$ or $P_g(s)$, should be more appropriate. The reason why our discussion was based on the Weibull and gamma distributions was that they proved to be in very good agreement with the probability density distributions of fragment size obtained by the theory of brittle fracture of one-dimensional ring-shaped materials, although no theoretical ground was explicitly provided in the literature\cite{Grady1981,Grady2008}. Following this existing study, we hypothesized that the similar distribution functions would be effective for two-dimensional brittle fracture, and eventually verified it experimentally using muskmelon rinds. Considering this, our findings are within a realm of ``empirical", instead of theoretical, as we presented in the title of the manuscript.

%----------------
\subsection{Practical utility of fragmentation law}

In practice, the quality of muskmelons is mainly evaluated by their appearance. In addition, the price difference due to the grade difference was much larger than that of the other fruits. Currently, this sorting work is performed visually by experts, such as producers and market participants. As far as the visual recognition range is concerned, muskmelons that are evaluated to be not for sale because of their low sugar content often have a non-fine netting pattern and irregular cell area distribution. However, relying on such craftsmanship can lead to arbitrary subjectivity in determining the muskmelon grades. Moreover, the number of muskmelons that can be identified manually is limited. If the relationship between the fruit growth process and the accompanying net formation process is clarified, it is possible to obtain an index that enables automatic and objective determination of the ripening degree of the fruit by visual inspection. Such an index can be applied to the cultivation management of muskmelons.

%----------------
\section{Conclusion}

We unveiled a hidden universal law that governs muskmelon rind fragmentation. The universal law is described by two types of probability distribution function, both of which are relevant to the modified Bessel function. The functional form of each distribution function can be derived by considering the expansion fracture of a complete spherical shell under internal pressure. These findings may open an avenue for the development of non-destructive diagnostic technology for muskmelons based on image analysis.

%----------------
\section*{Acknowledgment}
This work was supported by JSPS KAKENHI Grant Numbers 
JP21H00362, JP20J10344, JP19K03766, and JP18H03818.

\bibliographystyle{jpsj}
\bibliography{manu_Melon_JPSJ_ForArXiv}

\end{document}